# Networked buffering: a basic mechanism for distributed robustness in complex adaptive systems

**James M Whitacre[a]\* and Axel Bender[b]**

[a] School of Computer Science, University of Birmingham, Edgbaston, UK
[b] Land Operations Division, Defence Science and Technology Organisation; Edinburgh, Australia

\* Corresponding author

E-mail:
JW: j.m.whitacre@cs.bham.ac.uk
AB: axel.bender@dsto.defence.gov.au

# Abstract

A generic mechanism – *networked buffering* – is proposed for the generation of robust traits in complex systems. It requires two basic conditions to be satisfied: 1) agents are versatile enough to perform more than one single functional role within a system and 2) agents are degenerate, i.e. there exists partial overlap in the functional capabilities of agents. Given these prerequisites, degenerate systems can readily produce a distributed systemic response to local perturbations. Reciprocally, excess resources related to a single function can indirectly support multiple unrelated functions within a degenerate system. In models of genome:proteome mappings for which localized decision-making and modularity of genetic functions are assumed, we verify that such distributed compensatory effects cause enhanced robustness of system traits. The conditions needed for networked buffering to occur are neither demanding nor rare, supporting the conjecture that degeneracy may fundamentally underpin distributed robustness within several biotic and abiotic systems. For instance, networked buffering offers new insights into systems engineering and planning activities that occur under high uncertainty. It may also help explain recent developments in understanding the origins of resilience within complex ecosystems.

# Introduction

Robustness reflects the ability of a system to maintain functionality or some measured output as it is exposed to a variety of external environments or internal conditions. Robustness is observed whenever there exists a sufficient repertoire of actions to counter perturbations [1] and when a system's memory, goals, or organizational/structural bias can elicit those responses that match or counteract particular perturbations, e.g. see [2]. In many of the complex adaptive systems (CAS) discussed in this paper, the actions of agents that make up the system are based on interactions with a local environment, making these two requirements for robust behavior interrelated. When robustness is observed in such CAS, we generally refer to the system as being self-organized, i.e. stable properties spontaneously emerge without invoking centralized routines for matching actions and circumstances.

Many mechanisms that lead to robust properties have been distilled from the myriad contexts in which CAS, and particularly biological systems, are found [3] [4] [5] [6] [7] [8] [9] [10] [11] [12] [13] [14] [15] [16] [17] [18] [19] [20] [21]. For instance, robustness can form from loosely coupled feedback motifs in gene regulatory networks, from saturation effects that occur at high levels of flux in metabolic reactions, from spatial and temporal modularity in protein folding, from the functional redundancy in genes and metabolic pathways [22] [23], and from the stochasticity of dynamics[i] occurring during multi-cellular development [24] or within a single cell's interactome [25].

Although the mechanisms that lead to robustness are numerous and diverse, subtle commonalities can be found. Many mechanisms that contribute to stability act by responding to perturbations through local competitive interactions that appear cooperative at a higher level. A system's actions are rarely deterministically bijective (i.e. characterized by a one-to-one mapping between perturbation and response) and instead proceed through a concurrent stochastic process that in some circumstances is described as exploratory behavior [26].

This paper proposes a new basic mechanism that can lead to both local and distributed robustness in CAS. It results from a partial competition amongst system components and shares similarities with several of the mechanisms we have just mentioned. In the following, we speculate that this previously unexplored form of robustness may readily emerge within many different systems comprising multi-functional agents and may afford new insights into the exceptional flexibility that is observed within some complex adaptive systems.

In the next section we summarize accepted views of how diversity and degeneracy can contribute to robustness of system traits. We then present a mechanism that describes how a system of degenerate agents can create a widespread and comprehensive response to perturbations – the networked buffering hypothesis (Section 0). In Section 0 we provide evidence for the realisation of this hypothesis. We particularly describe the results of simulations that demonstrate that distributed robustness emerges from networked buffering in models of genome:proteome mappings. In Section 0 we discuss the importance of this type of buffering in natural and human-made CAS, before we conclude in Section 0. Three appendices supplement the content of the main body of this paper. In Appendix 1 we provide some detailed definitions for (and discriminations of) the concepts of degeneracy, redundancy and partial redundancy; in Appendix 2 we give background materials on degeneracy in biotic and abiotic systems; and in Appendix 3 we provide a technical description of the genome:proteome model that is used in our experiments.

# Robustness through Diversity and Degeneracy

As described by Holland [27], a CAS is a network of spatially distributed agents which respond concurrently to the actions of others. Agents may represent cells, species, individuals, firms, nations, etc. They can perform particular functions and make some of their resources (physical assets, knowledge, services, etc) work for the system.[ii] The control of a CAS tends to be largely decentralized. Coherent behavior in the system generally arises from competition and cooperation between agents; thus, system traits or properties are typically the result of the interplay between many individual agents.

Degeneracy refers to conditions where multi-functional CAS agents share similarities in only some of their functions. This means there are conditions where two agents can compensate for each other, e.g. by making the same resources available to the system, or can replace each other with regard to a specific function they both can perform. However, there are also conditions where the same agents can do neither. Although degeneracy has at times been described as partial redundancy, we distinctly differentiate between these two concepts. Partial redundancy only emphasizes the many-to-one mapping between components and functions while degeneracy concerns many-to-many mappings. Degeneracy is thus a combination of both partial redundancy and functional plasticity (explained below). We discuss the differences of the various concepts surrounding redundancy and degeneracy in Appendix 1 and Figure 1.

On the surface, having similarities in the functions of agents provides robustness through a process that is intuitive and simple to understand. In particular, if there are many agents in a system that perform a particular service then the loss of one agent can be offset by others. The advantage of having diversity amongst functionally similar agents is also straightforward to see. If agents are somewhat different, they also have somewhat different weaknesses: a perturbation or attack on the system is less likely to present a risk to all agents at once. This reasoning reflects common perceptions about the value of diversity in many contexts where CAS are found. For instance, it is analogous to what is described as *functional redundancy* [28] [29] (or response diversity [30]) in ecosystems, it reflects the rationale behind portfolio theory in economics and biodiversity management [31] [32] [33], and it is conceptually similar to the advantages from ensemble approaches in machine learning or the use of diverse problem solvers in decision making [34]. In short, diversity is commonly viewed as advantageous because it can help a system to consistently reach and sustain desirable settings for a single system property by providing multiple distinct paths to a particular state. In accordance with this thinking, examples from many biological contexts have been given that illustrate degeneracy's positive influence on the stability of a single trait, e.g. see Appendix 2. Although this view of diversity is conceptually and practically useful, it is also simplistic and, so we believe, insufficient for understanding how common types of diversity such as degeneracy will influence the robustness of multiple interdependent system traits.

CAS are frequently made up of agents that influence the stability of more than just a single trait because of their having a repertoire of functional capabilities. For instance, gene products act as versatile building blocks that form complexes with many distinct targets [35] [36] [37]. These complexes often have unique and non-trivial consequences inside or outside the cell. In the immune system, each antigen receptor can bind with (i.e. recognize) many different ligands and each antigen is recognized by many receptors [38] [39]; a feature that has only recently been integrated into artificial immune system models, e.g. [40] [41] [42]. In gene regulation, each transcription factor can influence the expression of several different genes with distinct phenotypic effects. Within an entirely different domain, people in organizations are versatile in the sense that they can take on distinct roles depending on who they are collaborating with and the current challenges confronting their team. More generally, the function an agent performs often depends on the context in which it finds itself. By context, we are referring to the internal states of an agent and the demands or constraints placed on the agent by its environment. As illustrated further in Appendix 2, this contextual nature of an agent's function is a common feature of many biotic and abiotic systems and it is referred to hereafter as *functional plasticity*.

Because agents are generally limited in the number of functions they are able to perform over a period of time, tradeoffs naturally arise in the functions an agent performs in practice. These tradeoffs represent one of several causes of trait interdependence and they obscure the process by which diverse agents influence the stability of single traits. A second complicating factor is the ubiquitous presence of degeneracy. While one of an agent's functions may overlap with a particular set of agents in the system, another of its functions may overlap with an entirely distinct set of agents. Thus functionally related agents can have additional compensatory effects that are differentially related to other agents in the system, as we describe in more detail in the next section. The resulting web of conditionally related compensatory effects further complicates the ways in which diverse agents contribute to the stability of individual traits with subsequent effects on overall system robustness.

# Networked Buffering Hypothesis

Previous authors discussing the relationship between degeneracy and robustness have described how an agent can compensate for the absence or malfunctioning of another agent with a similar function and thereby help to stabilize a single system trait. One aim of this paper is to show that when degeneracy is observed within a system, a focus on single trait robustness can turn away attention from a form of system robustness that spontaneously emerges as a result of a concurrent, distributed response involving chains of mutually degenerate agents. We organize these arguments around what we call the

*networked buffering hypothesis* (NBH). The central concepts of our hypothesis are described by referring to the abstract depictions of Figure 2; however, the phenomenon itself is not limited to these modeling conditions as will be elucidated in Section 0.

Consider a system comprising a set of multi-functional agents. Each agent performs a finite number of tasks where the types of tasks performed are constrained by an agent's functional capabilities and by the environmental requirement for tasks ("requests"). A system's robustness is characterized by the ability to satisfy tasks under a variety of conditions. A new "condition" might bring about the failure or malfunctioning of some agents or a change in the spectrum of environmental requests. When a system has many agents that perform the same task then the loss of one agent can be compensated for by others, as can variations in the demands for that task. Stated differently, having an excess of functionally similar agents (excess system resources) can provide a buffer against variations in task requests.

In the diagrams of Figure 2, for sake of illustration the multi-functionality of CAS agents is depicted in an abstract "functions space". In this space, bi-functional agents (represented by pairs of connected nodes) form a network (of tasks or functions) with each node representing a task capability. The task that an agent currently performs is indicated by a dark node, while a task that is not actively performed is represented by a light node. Nodes are grouped into clusters to indicate functional similarity amongst agents. For instance, agents with nodes occupying the same cluster are said to be similar with respect to that task type. To be clear, task similarity implies that either agent can adequately perform a task of that type making them interchangeable with respect to that task. In Figure 2d we illustrate what we call 'pure redundancy' or simply 'redundancy': purely redundant agents are always functionally identical in either neither or across both of the task types they can perform. In all other panels of Figure 2, we show what we call 'pure degeneracy': purely degenerate agents either cannot compensate for each other or can do so in only one of the two task types they each can carry out..

Important differences in both scale and the mechanisms for achieving robustness can be expected between the degenerate and redundant system classes. As shown in Figure 2b, if more (agent) resources are needed in the bottom task group and excess resources are available in the top task group, then degeneracy allows agents to be reallocated from tasks where they are in excess to tasks where they are needed. This occurs through a sequence of reassignments triggered by a change in environmental conditions (as shown in Figure 2b by the large arrows with switch symbols) – a process that is autonomous so long as agents are driven to complete unfulfilled tasks matching their functional repertoire.

Figure 2b illustrates a basic process by which resources related to one type of function can support unrelated functions. This is an easily recognizable process that can occur in each of the different systems that are listed in Table 1. In fact, conditional interoperability is so common within some domains that many domain experts would consider this an entirely unremarkable feature. What is not commonly appreciated though is that the number of distinct paths by which reconfiguration of resources is possible can potentially be enormous in highly degenerate systems, depending on where resources are needed and where they are in excess (see Figure 2c). Conversely, this implies that it is theoretically possible for excess agent resources (buffers) in one task to indirectly support an enormous number of other tasks, thereby increasing the effective versatility of any single buffer (seen if we reversed the flow of reassignments in Figure 2c). Moreover, because buffers in a degenerate system are partially related, the stability of any system trait is potentially the result of a distributed, networked response within the system. For instance, resource availability can arise through an aggregated response from several of the paths shown in Figure 2c. Although interoperability of agents may be localized, extra resources can offer huge reconfiguration opportunities at the system level.

These basic attributes are not feasible in reductionist systems composed of purely redundant agents (Figure 2d). Without any partial overlap in capabilities, agents in the same functional groups can only support each other and, conversely, excess resources cannot support unrelated tasks outside the group. Buffers are thus localized. In the particular example illustrated in Figure 2d, agent resources are always tied to one of two types of tasks. Although this ensures certain levels of resources will always remain available within a given group, it also means they are far less likely to be utilized when resource requirements vary, thereby reducing resource efficiency. In other words, resource buffers in purely redundant systems are isolated from each other, limiting how versatile the system can be in reconfiguring these resources. In fact, every type of variability in task requirements needs a matching realization of redundancies. If broad reconfigurations are required (e.g. due to a volatile environment) then these limitations will adversely affect system robustness. Although such statements are not surprising, they are not trivial either because the sum of agent capabilities within the redundant and degenerate systems are identical.

## Networked Buffering in Genome: Proteome Mappings

More than half of all mutational robustness in genes is believed to be the result of distributed actions and not genetic redundancy [4]. Although a similar analysis of the origins of robustness has not taken place for other biotic contexts, there is plenty of anecdotal evidence for the prevalence of both local functional redundancy *and* distributed forms of robustness in biology. Degeneracy may be an important causal factor for both of these forms of robustness. Edelman and Gally have presented considerable evidence of degeneracy's positive influence on functional redundancy, i.e. single trait stability through *localized* compensatory actions, see [23], Section 0 and Appendices 1 and 2. What is missing though is substantiation for degeneracy's capacity to cause systemic forms of robustness through *distributed* compensatory actions.

In the previous section we hypothesized how degeneracy might elicit distributed robustness through networked sequences of functional reassignments and resource reconfigurations. To substantiate this hypothesis, we evaluate robustness in a model of genome:proteome (G:P) mappings that was first studied in [43]. In the model, systems of proteins ("agents") are driven to satisfy environmental conditions through the utilization of their proteins. Protein-encoding genes express a single protein. Each protein has two regions that allow it to form complexes with ligands that have a strong affinity to those regions (see Figure 3). A protein's "behavior" is determined by how much time it spends interacting with each of the target ligands. The sum of protein behaviors defines the system phenotype, assuming that each protein's trait contributions are additive. It is further assumed that genetic functions are modular [44] such that there are little or no restrictions in what types of functions can be co-expressed in a single gene or represented in a single protein.[iii] The environment is defined by the ligands available for complex formation. Each protein is presented with the same well-mixed concentrations of ligands. A phenotype that has unused proteins is energetically penalized and is considered unfit when the penalty exceeds a predefined threshold. Two types of systems are evaluated: those where the G:P mapping is purely redundant (as of the abstract representation in Figure 2d) and those where it is purely degenerate (as of Figure 2a). For more details on the model see [43] and Appendix 3.

In [43], we found that purely degenerate systems are more robust to perturbations in environmental conditions than are purely redundant ones, with the difference becoming larger as the systems are subjected to increasingly larger perturbations (Figure 4a). In addition we measured the number of distinct null mutation combinations under which a system could maintain fitness and found that degenerate systems are also much more robust with respect to this measurement ("versatility") [43]. Importantly, this robustness improvement becomes more pronounced as the size of the systems increases (Figure 4b).

We now expand on the studies of [43] by showing that the enhanced robustness in purely degenerate systems originates from distributed compensatory effects. First, in Figure 4d we repeat the experiments used to evaluate system versatility; however, we restrict the systems' response options to local actions only. More precisely, only proteins of genes that share some functional similarity to the products of the mutated genes are permitted to change their behaviors and thus participate in the system's response to gene mutations. By adding this constraint to the simulation, the possibility that distributed compensatory pathways (as described in Figure 2b and c) can be active is eliminated. In other words, this constraint allows us to measure the robustness that results from direct functional compensation; i.e. the type of robustness in those examples of the literature where degeneracy has been related to trait stability, e.g. see [23].

In Figure 4d the robustness of the purely redundant systems remains unchanged compared with the results in Figure 4b while the robustness of degenerate systems degrades to values that are indistinguishable from the redundant system results. Comparing the two sets of experiments, we find that roughly half of the total robustness that is observable in the degenerate G:P models originates from non-local effects that cannot be accounted for by the relationships between degeneracy and robustness that were previously described in the literature, e.g. in [23].

As further evidence of distributed robustness in degenerate G:P mappings, we use the same conditions as in Figure 4b except now we systematically introduce single loss of function mutations and record the proportion C of distinct gene products that change state. In the probability distributions of Figure 4c, the redundant systems only display localized responses as would be expected while the degenerate systems respond to a disturbance with both small and large numbers of changes to distinct gene products.

As small amounts of excess resources are added to degenerate systems (Figure 5a), single null mutations tend to invoke responses in a larger number of distinct gene products while robustly maintaining system traits, i.e. system responses become more distributed while remaining phenotypically cryptic. In measuring the magnitude S of state changes for individual gene products, we find the vast majority of state changes that occur are consistently small across experiments (making them hard to detect in practice), although larger state changes become more likely when excess resources are introduced (Figure 5b). The effect from adding excess resources saturates quickly and shows little additional influence on system properties (C and S) for excess resources > 2%.

Individually varying other parameters of the model such as the maximum rate of gene expression, the size of the genetic system, or the level of gene multi-functionality did not alter the basic findings reported here. Thus for the degenerate models of G:P mappings, we find that distributed responses play an important role in conferring mutational robustness towards single null mutations. Although our experimental conditions differ in some respects from the analysis of single gene knockouts in *Saccharomyces cerevisiae* [4], both our study and [4] find evidence that roughly half the mutational robustness of genetic systems is a consequence of distributed effects: a finding that is similar to observations of robustness in the more specific case of metabolic networks [13].

The robustness we evaluate in our experiments only considers loss of function mutations. However, we experimentally observed similar relationships between degeneracy and distributed robustness when we expose our model systems to small environmental perturbations, e.g. changes to ligand concentrations. This is suggestive not only of congruency in the robustness towards distinct classes of perturbations [45], but also that distributed robustness is conferred in this model through the same mechanistic process, i.e. a common source of biological canalization as proposed in [46]. As supported by the findings of other studies [14] [43] [47] [48], the observation of an additional "emergent form" of robustness also suggests that robustness is neither a conserved property of complex systems nor does it have a conceptually intuitive trade-off with resource efficiency as has been proposed in discussions related to the theory of Highly Optimized Tolerance [3] [7] [49].

## Discussion

There is a long-standing interest in the origins of robustness and resilience within CAS in general and biological systems in particular [28] [45] [46] [48] [49] [50] [51] [52] [53] [54] [55] [56] [57] [58] [59]. Although considerable progress has been made in understanding constraint/deconstraint processes in biology [26], a full account of biological robustness remains elusive. The extent to which degeneracy can fill this knowledge gap is unknown, however we outline several reasons why degeneracy might play a vital role in facilitating local and distributed forms of biological robustness.

### *Omnipresence of degeneracy in biological CAS*

Stability under *moderately* variable conditions (i.e. modest internal or external changes to the system) is a defining attribute of biology at all scales [46] [60].[iv] Any mechanism that broadly contributes to such stability must be as ubiquitous as the robust traits it accounts for. Although many mechanisms studied in the literature (such as those mentioned in the Introduction) are broadly observed, few are as pervasive as degeneracy. In fact, degeneracy is readily seen throughout molecular, genetic, cellular, and population levels in biology [23] and it is a defining attribute of many communication and signalling systems in the body including those involved in development, immunity, and the nervous system [23] [38] [61] [62]. As described in Appendix 2, degeneracy is also readily observed in other complex adaptive systems including human organizations, complex systems engineering, and ecosystems. When the degenerate components of these systems form a network of partially overlapping functions, and when component responses are fast relative to the timescale of perturbations, we argue that networked buffering should, in principle, enhance the robustness and flexibility observed in each of these distinct system classes.

### *Cellular robustness*

If degeneracy broadly accounts for biological robustness then it should be intimately related to many mechanisms discussed in the literature. One prominent example where this occurs is the relationship between degeneracy and cell regulation. For example, the organization or structure of metabolic reactions, signalling networks, and gene expression elicits some control over the sequences of interactions that occur in the 'omic' network. This control is often enacted by either a process of competitive exclusion or recruitment within one of the interaction steps in a pathway (e.g. a metabolite in a reaction pathway or the initial binding of RNA polymerase prior to gene transcription).

Given the reductionist bias in science, it is not surprising that biologists initially expected to find a single molecular species for every regulatory action. Today however most of the accumulated evidence indicates that local regulatory effects are often enacted by a number of compounds that are degenerate in their affinity to particular molecular species and are roughly interchangeable in their ability to up/down regulate a particular pathway. NBH suggests that when the relationships between degenerate regulators form a network of partial competition for regulatory sites, this may confer high levels of regulatory stability, e.g. against stochastic fluctuations for stabilizing gene expression [63] or, and more generally, towards more persistent changes in the concentrations of molecular species.

On the other hand, when degeneracy is absent then the regulatory processes in biology are more sensitive to genetic and environmental perturbations, although in some case this sensitivity is useful, e.g. in conferring stability to traits at a higher level. However in the complete absence of degeneracy, only one type of molecular species could be responsible for each type of control action, and the removal of that species could not be directly compensated for by others. Under these conditions, change in function mutations to non-redundant genes would most likely result in changes to one or more traits. In other words, mutational robustness would be greatly reduced and cryptic genetic variation would not be observed in natural populations.

### *Systems engineering*

The redundancy model in Figure 2d reflects a logical decomposition of a system that is encouraged (though not fully realized) in most human planning/design activities, e.g. [64] [65]. While there are many circumstances where redundancy is beneficial, there are others where we now anticipate it will be detrimental. Redundancy can afford economies of scale and provide transparency, which can allow a system to be more amenable to manipulation by bounded-rational managers (cf

[66] [67] [68]). When systems or subsystems operate within a predictable environment with few degrees of freedom, redundancy/decomposition design principles have proven to be efficient *and* effective. However, when variability in conditions is hard to predict and occurs unexpectedly, purely redundant and decomposable system architectures may not provide sub-systems with the flexibility necessary to adapt and prevent larger systemic failures. Under these circumstances, we propose that networked buffering from degeneracy can improve system stability. We are currently involved in a project that is exploring these ideas in the context of algorithm design and strategic planning and we have now accumulated some evidence that networked buffering is a relevant attribute for some systems engineering contexts [47] [69] [70].

### *Weak Links in complex networks*

Within fluid markets, social systems, and each of the examples listed in Table 1, one can find systems composed of functionally plastic degenerate components that operate within a dynamic uncertain world. We have argued that the ability of these components to partially overlap across different function classes will lead to the emergence of networked buffering. This functional compensation, however, is not always easy to detect. When agents are functionally plastic, they tend to interact with many distinct component types. This behavior causes individual interaction strengths to appear weak when they are evaluated in aggregation using time-averaged measurements, e.g. see [71] and Figure 5b. As we elaborate in [72], commonly accepted forms of experimental bias tend to overlook weak interactions in the characterization and analysis of CAS networks. Yet there is a growing number of examples (e.g. in social networks and proteomes) where weak links contribute substantially to system robustness as well as similar properties such as system coherence [73] [74]. Particularly in the case of social networks, degenerate weak links help to establish communication channels amongst cliques and support cohesion within the social fabric through processes that mirror the basic principles outlined in NBH, e.g. see [73] [74].

### *Ecosystem Resilience*

In a world undergoing regional environmental regime shifts brought about by changes in the global climate, it is becoming increasingly important to understand what enables ecosystems to be resilient, i.e. to tolerate disturbances without shifting into qualitatively different states controlled by different sets of processes [29]. Ecology theory and decades of simulation experiments have concluded that increasing complexity (increasing numbers of species and species interactions) should destabilize an ecosystem. However, empirical evidence suggests that complexity and robustness are positively correlated. In a breakthrough study, Kondoh [75] [76] has demonstrated that this paradox can be resolved within biologically plausible model settings when two general conditions are observed: i) species are functionally plastic in resource consumption (adaptive foraging) and ii) potential connectivity in the food web is high. Because higher connectivity between functionally plastic species allows for degeneracy to arise, Kondoh's requirements act to satisfy the two conditions we have set out for the emergence of networked buffering and its subsequent augmenting of system stability. We therefore advocate that the findings of [75] may provide the first direct evidence that degeneracy and networked buffering are necessary for positive robustness–complexity relationships to arise in ecosystems. Other recent studies confirm that including degeneracy within ecosystem models results in unexpected non-localized communication in ecosystem dynamics [77]. We propose that these non-local effects could be another example of the basic resource rearrangement properties that arise due to networked buffering.

Despite rich domain differences, we contend there are similarities in how the organizational properties in several CAS facilitate flexibility and resilience within a volatile environment. While the potential advantages from networked buffering are obvious, our intention here is not to make claims that it is the only mechanism that explains the emergence of robustness in system traits. Nor is it our intent to make general claims about the adaptive significance of this robustness or to imply selectionist explanations for the ubiquity of degeneracy within the systems discussed in this article. Much degeneracy is likely to be passively acquired in nature (e.g. see [72]). Moreover, there are instances where trait stability is not beneficial as is illustrated in [78] [79] [80] where examples of mal-adaptive robustness in biological and abiotic contexts is provided.

## Conclusions

This paper introduces what is argued to be a new mechanism for generating robustness in complex adaptive systems that arises due to a partial overlap in the functional roles of multi-functional agents; a system property also known in biology as degeneracy. There are many biological examples where degeneracy is already known to provide robustness through the local actions of functionally redundant components. Here however we have presented a conceptual model showing how degenerate agents can readily form a buffering network whereby agents can indirectly support many functionally dissimilar tasks. These distributed compensatory effects result in greater versatility and robustness – two characteristics with obvious relevance to systems operating in highly variable environments.

Recent studies of genome-proteome models have found degenerate systems to be exceptionally robust in comparison to those without degeneracy. Expanding on these results, we have tested some of the claims of the buffering network hypothesis and determined that the enhanced robustness within these degenerate genome:proteome mappings is in fact a consequence of distributed (non-local) compensatory effects that are not observable when robustness is achieved using only pure redundancy. Moreover, the proportion of local versus non-local sources of robustness within the degenerate models shows little sensitivity to scaling and is compatible with biological data on mutational robustness.

**Author's Contributions**: JW designed and carried out experiments. JW and AB wrote the paper and interpreted the results. Both authors have read and approved the final manuscript.

**Acknowledgements:** This research was partially supported by DSTO and an EPSRC grant (No. EP/E058884/1).

**Competing Interests:** The authors declare that they have no competing interests

## Appendix 1: Degeneracy, Redundancy, and Partial Redundancy

Redundancy and degeneracy are two system properties that contribute to the robustness of biological systems [4] [23]. Redundancy is an easily recognizable property that is prevalent in both biological and man-made systems. Here, redundancy means 'redundancy of parts' and refers to the coexistence of identical components with identical functionality (i.e. the components are isomorphic and isofunctional). In information theory, redundancy refers to the repetition of messages, which is important for reducing transmission errors. Redundancy is also a common feature of engineered or planned systems where it provides robustness against variations of a very specific type ('more of the same' variations). For example, redundant parts can substitute for others that malfunction or fail, or augment output when demand for a particular output increases.

Degeneracy differs from pure redundancy because similarities in the functional response of components are not observed for all conditions (see Figure 1d). In the literature, degeneracy has at times been referred to as functional redundancy or partial redundancy, however most definitions for these terms only emphasize the many-to-one mapping between components and functions (e.g. [9] [81] [82] [83] [84] [85] [86]). On the other hand, the definition of degeneracy used here and in [23] [77] [87] [88] [89] [90] also emphasizes a one-to-many mapping.

To put it more distinctly, our definition of degeneracy requires degenerate components to also be functionally versatile (one-to-many mapping), with the function performed at any given time being dependent on the context; a behavior we label as *functional plasticity* [77] [90]. For degeneracy to be present, some (but not all) functions related to a component or module must also be observable in others, i.e. a partial and conditional similarity in the repertoire of functional responses (see Figure 1). In contrast, partial redundancy is often used to describe the conditional similarity in functional responses for components capable of only a single function (see Figure 1c). This is analogous to the definition of response diversity within ecosystems [30][v] and is conceptually similar to ensemble approaches in machine learning.

Functional plasticity is necessary to create the buffering networks discussed in Section 0 and the enhanced evolvability observed in [43] [47]. However this requirement is not as demanding as it may at first seem. Functional plasticity is common in biological systems and occurs for most cited examples of degeneracy in [23]. For instance, gene products such as proteins typically act like versatile building blocks, performing different functions that depend on the complex a protein forms with other gene products or other targets in its environment [91] [92]. In contrast to earlier ideas that there was one gene for each trait, gene products are now know to have multiple non-trivial interactions with other "targets", i.e. in the interactome [36] [37] and these are rarely correlated in time [93]. The alternative, where a gene's functions are all performed within the same context (referred to as "party hubs" in [93]), is known to be considerably less common in biology.

## Appendix 2: Degeneracy in biotic and abiotic systems

In biology, degeneracy refers to conditions where the functions or capabilities of components overlap partially. In a review by Edelman and Gally [23], numerous examples are used to demonstrate the prevalence of degeneracy throughout biology. It is pervasive in proteins of every functional class (e.g. enzymatic, structural, or regulatory) [90] [94] and is readily observed in ontogenesis (see page 14 in [95]), the nervous system [87] and cell signalling (crosstalk). In the particular case of proteins, it is also now known that partial functional similarities can arise even without any obvious similarities in sequence or structure [96].

Degeneracy and associated properties like functional plasticity are also prevalent in other biotic and abiotic systems, such as those listed below in Table 1. For instance, in transportation fleets the vehicles are often interchangeable but only for certain tasks. Multi-functional force elements within a defence force structure also can exhibit an overlap in capabilities but only within certain missions or scenarios. In an organization, people often have overlapping job descriptions and are able to take on some functions that are not readily achieved by others that technically have the same job. In the food webs of complex ecosystems, species within similar trophic levels sometimes have a partial overlap in resource competition. Resource conditions ultimately determine whether competition will occur or whether the two species will forage for distinct resources [75].

Degeneracy has become increasingly appreciated for its role in trait stability, as was noted in [72] and more thoroughly discussed in [23]. For instance, gene families can encode for diverse proteins with many distinctive roles yet sometimes these proteins can compensate for each other during lost or suppressed gene expression, as seen in the developmental roles of the adhesins gene family in *Saccharomyces* [97]. At higher scales, resources are often metabolized by a number of distinct compensatory pathways that are effectively interchangeable for certain metabolites even though the total effects of each pathway are not identical.

More generally, when agents are degenerate some functions will overlap meaning that the influence an agent has in the system could alternatively be enacted by other agents, groups of agents, or pathways. This functional redundancy within a specified context provides the basis for both competition and collaboration amongst agents and in many circumstances can contribute to the stability of individual traits (cf. [23]).

## Appendix 3: Technical description of genome:proteome model

The genome:proteome model was originally developed in [43] and consists of a set of genetically specified proteins (i.e. material components). Protein state values indicate the functional targets they have interacted with and also define the trait values of the system. The genotype determines which traits a protein is able to influence, while a protein's state dictates how much a protein has actually contributed to each of the traits it is capable of influencing. The extent to which a protein $i$ contributes to a trait $j$ is indicated by the matrix elements $M_{ij} \in Z$. Each protein has its own unique set of genes, which are given by a set of binary values $\delta_{ij}$, $i \in n$, $j \in m$. The matrix element $\delta_{ij}$ takes a value of one if protein $i$ can functionally contribute to trait $j$ (i.e. bind to protein target $j$) and zero otherwise. In our experiments, each gene expresses a single protein (i.e. there is no alternative splicing). To simulate the limits of functional plasticity, each protein is restricted to contribute to at most two traits, i.e. $\sum_{i \in n} \delta_{ij} \leq 2 \; \forall i$. To model limits on protein utilization (e.g. as caused by the material basis of gene products), maximum trait contributions are defined for each protein, which for simplicity are set equal, i.e. $\sum_{j \in m} M_{ij} \delta_{ij} = \lambda$ $\forall i$ with the integer $\lambda$ being a model parameter.

The set of system traits defines the system phenotype with each trait calculated as a sum of the individual protein contributions $T_j^P = \sum_{i \in n} M_{ij} \delta_{ij}$. The environment is defined by the vector $T^E$, whose components stipulate the number of targets that are available. The phenotypic attractor $F$ is defined in Eq. 1 and acts to (energetically) penalize a system configuration when any targets are left in an unbound state, i.e. $T_j^P$ values fall below the satisfactory level $T_j^E$.

**Simulation:** Through control over its phenotype a system is driven to satisfy the environmental conditions. This involves control over protein utilization, i.e. the settings of $M$. We implement *ordered asynchronous updating* of $M$ where each protein stochastically samples local changes in its utilization (changes in state values $M_{ij}$ that alter the protein's contribution to system traits). Changes are kept if compatible with the global attractor for the phenotype defined by Eq. 1. Genetic mutations involve modifying the gene matrix $\delta$. For mutations that cause loss of gene function, we set $\delta_{ij} = 0 \; \forall j$ when gene $i$ is mutated.

$$F(T^P) = -\sum_{j \in m} \theta_j$$

$$\theta_j = \begin{cases} 0, & T_j^P > T_j^E \\ (T_j^P - T_j^E), & else \end{cases}$$

(1)

**Degenerate Systems:** We model degeneracy and redundancy by constraining the settings of the matrix $\delta$. This controls how the trait contributions of proteins are able to overlap. In the '*redundant model*', proteins are placed into subsets in which all proteins are genetically identical and thus influence the same set of traits. However, redundant proteins are free to take on distinct state values, which reflects the fact that proteins can take on different functional roles depending on their local context. In the '*degenerate model*', proteins can only have a partial overlap in what traits they are able to affect. The intersection of trait sets influenced by two degenerate proteins is non-empty and truly different to their union. An illustration of the redundant and degenerate models is given in Figure 3.

## Appendix: Notes

[i] Stochasticity enhances robustness but is not technically a mechanism for achieving it. Over time, stochasticity forces the states and structures of a system towards paths that are less sensitive to natural fluctuations and this provides "robustness for free" to any other congruent perturbations that were not previously observed.

[ii] In this sense, agents *are* resources. In the models presented in Figure 2, Section 0 and Appendix 3 we assume, without loss of generality, that agent resources are reusable.

[iii] In a forthcoming paper we provide evidence that the findings in [43] are typically not affected by constraints on the functional combinations allowed within a single gene.

[iv] Our emphasis on robustness towards small/moderate changes is an acknowledgement of the contingency of robustness that is observed in CAS, e.g. the niches of individual species. Mentioning robustness to different classes of perturbation is not meant to imply robustness measurements are not affected by the type of perturbation. Instead it reflects our belief that the mechanistic basis by which robustness is achieved is similar in both cases, i.e. there is a common cause of canalization [46].

[v] Response diversity is defined as the range of reactions to environmental change among species contributing to the same ecosystem function.

[vi] Note that the diagrams of redundant and degenerate systems represent educative examples only. In many biotic and abiotic CAS, agents are able to perform more than two functions. Also, in practice, systems with multi-functional agents will have some degree of both redundancy and degeneracy. For instance, if the circled agent in panel (a) were introduced to the system in panel (d) then that system would have partially overlapping buffers and thus some small degree of degeneracy.

# Tables

**Table 1: Systems where agents are multifunctional and have functions that can partially overlap with other agents.**

| Agent | System | Environment | Control | Agent Tasks |
|---|---|---|---|---|
| Vehicle type | Transportation Fleet | Transportation Network | Centralized Command and Control | Transporting goods, pax |
| Force element | Defence Force Structure | Future Scenarios | Strategic Planning | Missions |
| Person | Organization | Marketplace | Management | Job Roles |
| Deme | Ecosystem | Physical Environment | Self-organized | Resource usage and creation |
| Gene Product | Interactome | Cell | Self-organized and evolved | Energetic and sterric interactions |
| Antigen | Immune System | Antibodies and host proteins | Immune learning | Recognizing foreign proteins |

# Figures

**Figure 1** Illustration of degeneracy and related concepts. Components (C) within a system have a functionality that depends on their context (E) and can be functionally active (filled nodes) or inactive (clear nodes). When a component exhibits qualitatively different functions (indicated by node color) that depend on the context, we refer to that component as being functionally plastic (panel a). Pure redundancy occurs when two components have identical functions in every context (panels b and c). Functional redundancy is a term often used to describe two components with a single (but same) function whose activation (or capacity for utilization) depends on the context in different ways (panel d). Degeneracy describes components that are functionally plastic and functionally redundant, i.e. where the functions are similar in some situations but different in others (panel e).

**Figure 2:** Conceptual model of a buffering network. Each agent is depicted by a pair of connected nodes that represent two types of tasks/functions that the agent can perform, e.g. see dashed circle in panel a). Node pairs that originate or end in the same node cluster ("Functional group") correspond to agents that can carry out the same

function and thus are interchangeable for that function. Darkened nodes indicate the task an agent is currently performing. If that task is not needed then the agent is an excess resource or "buffer". Panel a) Degeneracy in multi-functional agents. Agents are degenerate when they are only similar in one type of task. Panel b) End state of a sequence of task reassignments or resource reconfigurations. A reassignment is indicated by a blue arrow with switch symbol. The diagram illustrates a scenario in which requests for tasks in the Z functional group have increased and requests for tasks of type X have decreased. Thus resources for X are now in excess. While no agent exists in the system that performs both Z and X, a pathway does exist for reassignment of resources (X→Y, Y→Z). This illustrates how excess resources for one type of function can indirectly support unrelated functions. Panel c) Depending on where excess resources are located, reconfiguration options are potentially large as indicated by the different reassignment pathways shown. Panel d) A reductionist system design with only redundant system buffers cannot support broad resource reconfiguration options. Instead, agent can only participate in system responses related to its two task type capabilities.[vi]

Figure 3: Overview of genome-proteome model. a) Genotype-phenotype mapping conditions and pleiotropy: Each gene contributes to system traits through the expression of a protein product that can bind with functionally relevant targets (based on genetically determined protein specificity). b) Phenotypic expression: Target availability is influenced by the environment and by competition with functionally redundant proteins. The attractor of the phenotype can be loosely described as the binding of each target with a protein. c) Functional overlap of genes: Redundant genes can affect the same traits in the same manner. Degenerate traits only have a partial similarity in what traits they affect.

Figure 4: Local and distributed sources of robustness in protein systems designed according to purely redundant and purely degenerate G:P mappings. a) Differential robustness as a function of the percentage of genes that are mutated in each protein system. Differential robustness is defined as the probability for a system phenotype to maintain fitness after it was allowed to adjust to a change in conditions (here: gene mutations). Source: [43] b) Versatility-robustness as a function of initial excess protein resources. Versatility is measured as the number of null mutation combinations ("neutral network size") for which the system phenotype maintains fitness. Source: [43]. c) Frequency distribution for the proportion C of distinct gene products that change their function when versatility is evaluated (as of panel b experiments) in systems with 0% initial excess resources. d) Versatility of redundant and degenerate systems when the system response to null mutations is restricted to local compensation only; i.e. gene products can only change their functional contribution if they are directly related to those functions lost as a result of a null mutation.

Figure 5 Probability distributions for a) proportion C of distinct gene products that change state and b) magnitude S of change in gene products. Experiments are shown for degenerate G:P mappings using the same conditions as in Figure 4b, but with the following modifications: 1) perturbations to the system are of single null mutations only and 2) systems are initialized with different amounts of excess resources (% excess indicated by data set label).

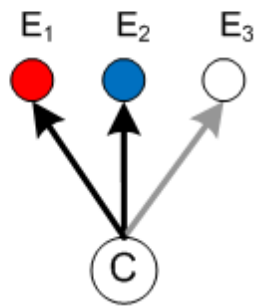
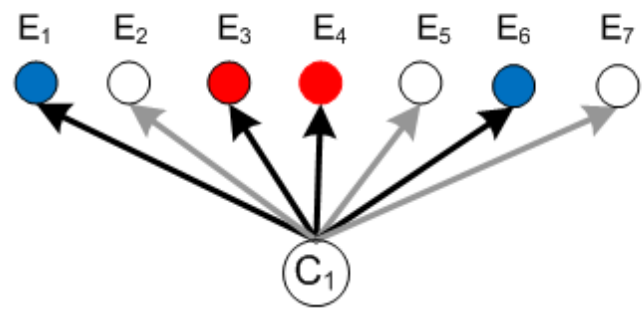
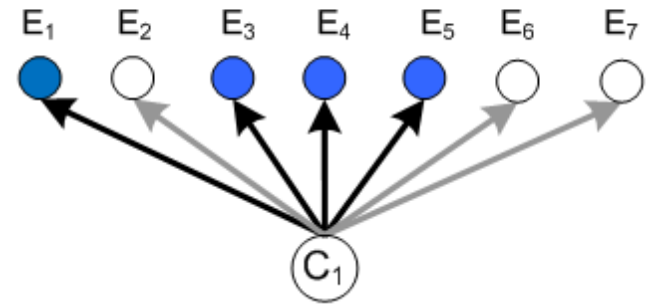
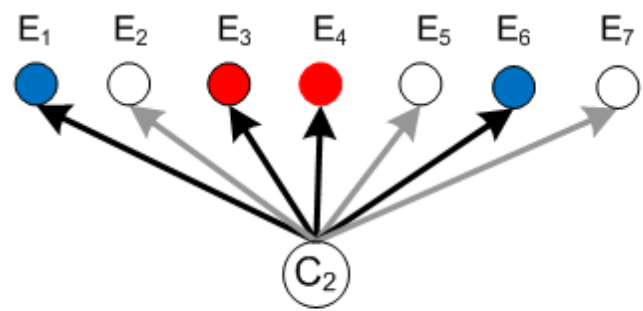
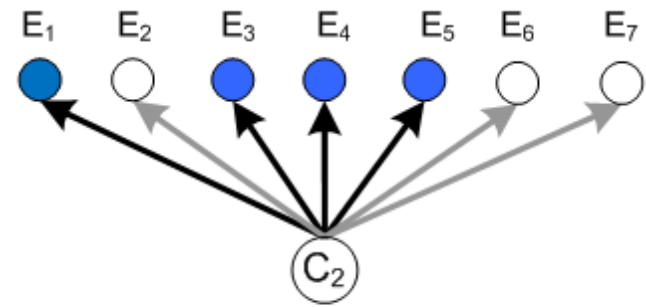
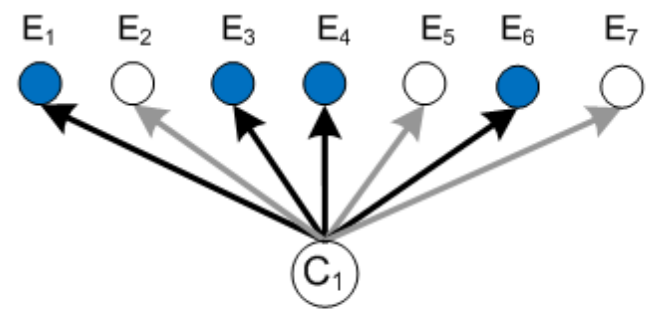
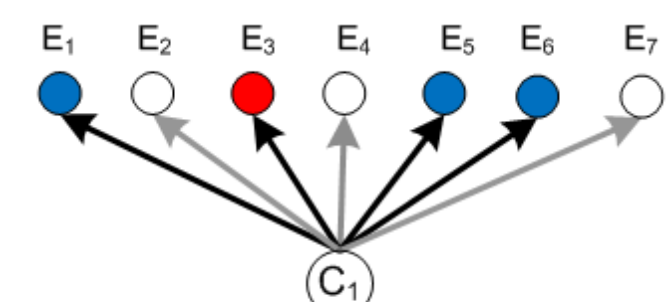
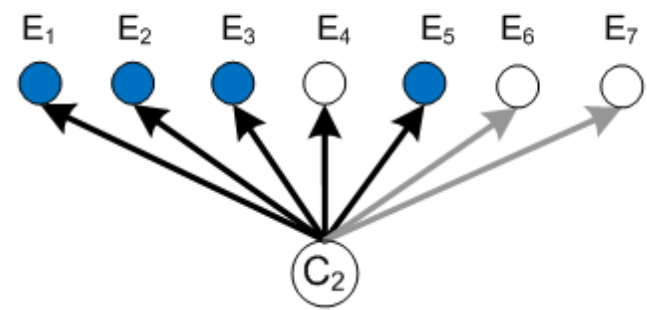
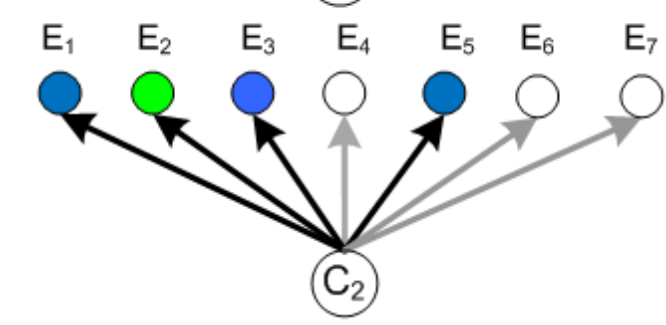

Figure 1

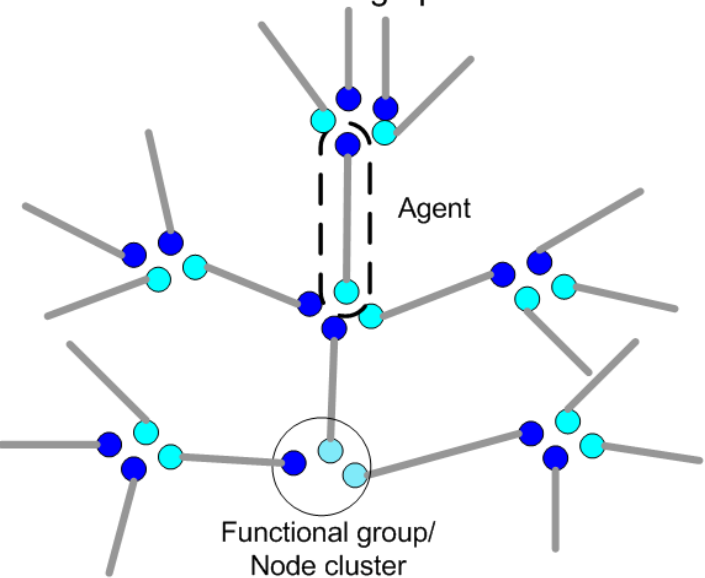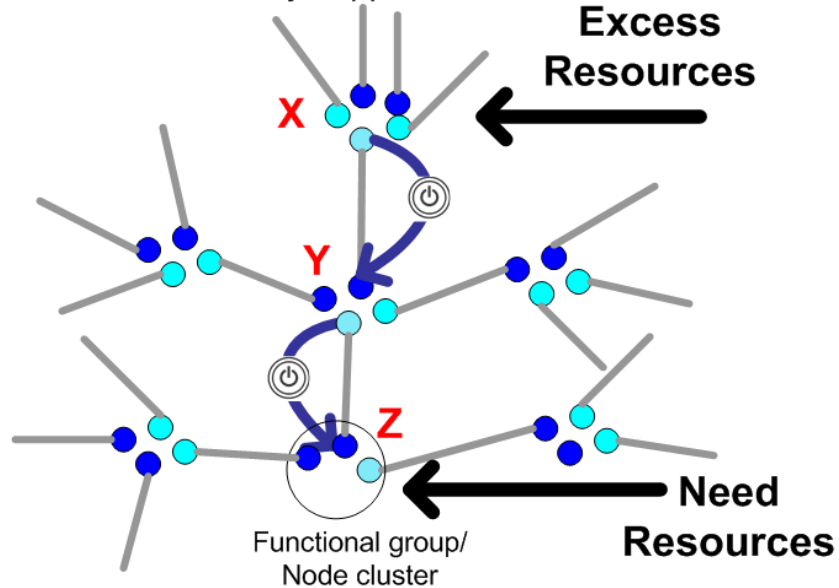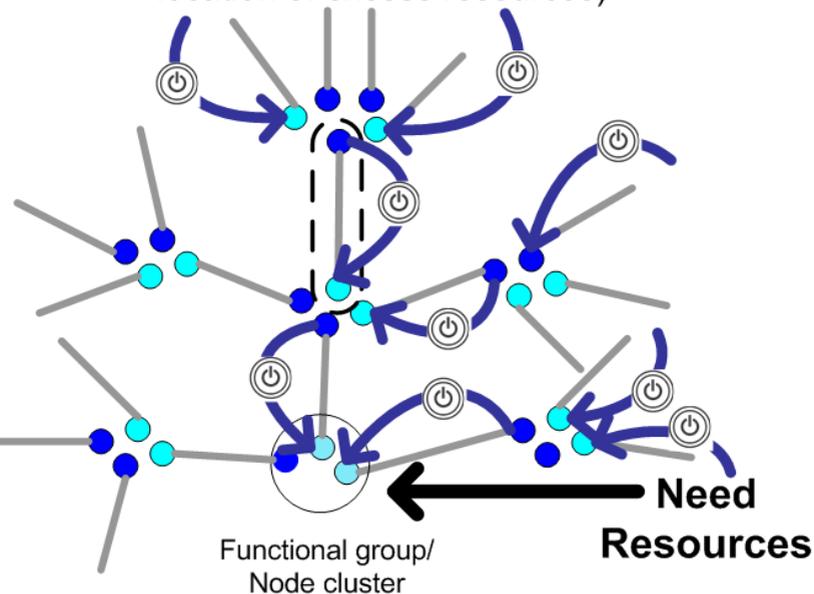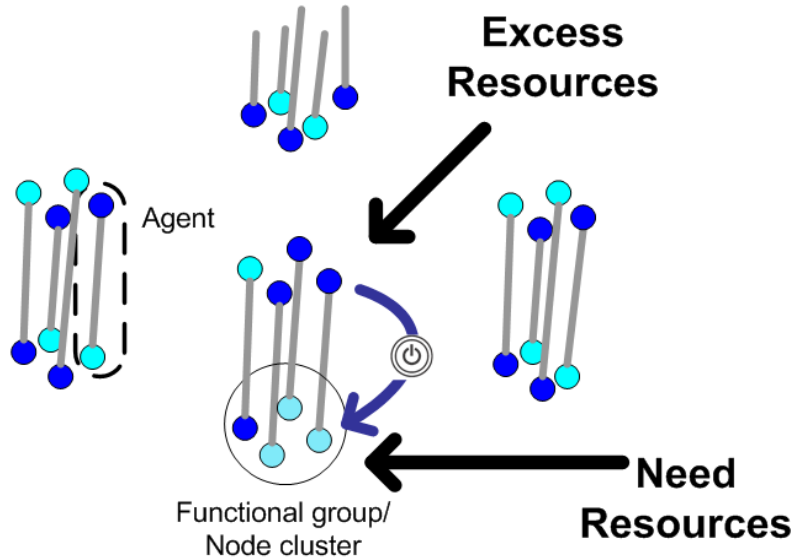

Figure 2

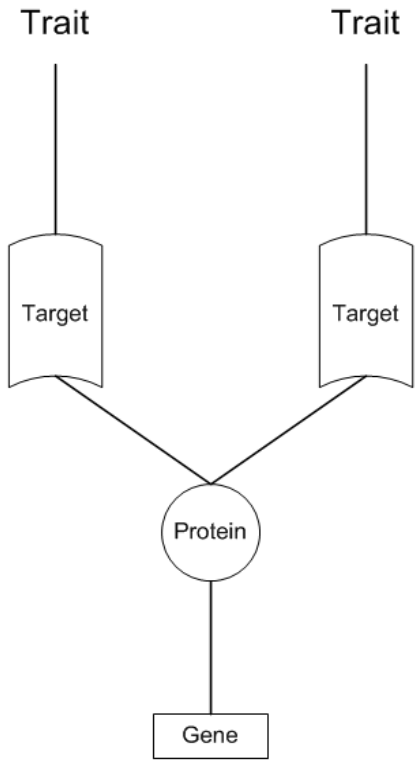
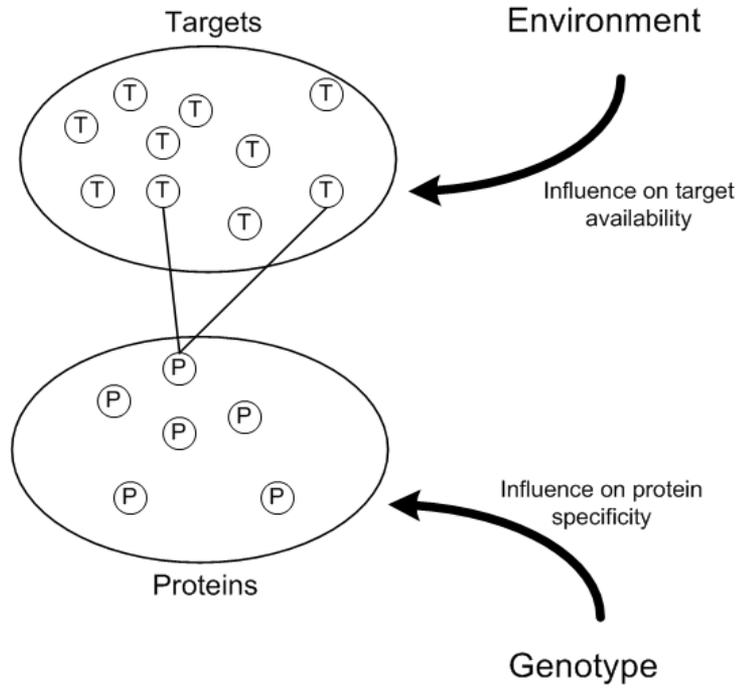
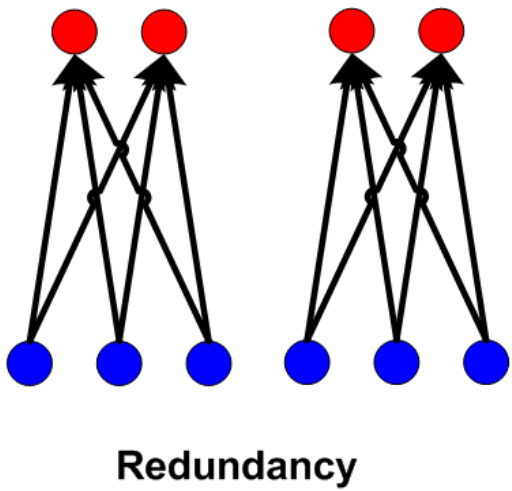
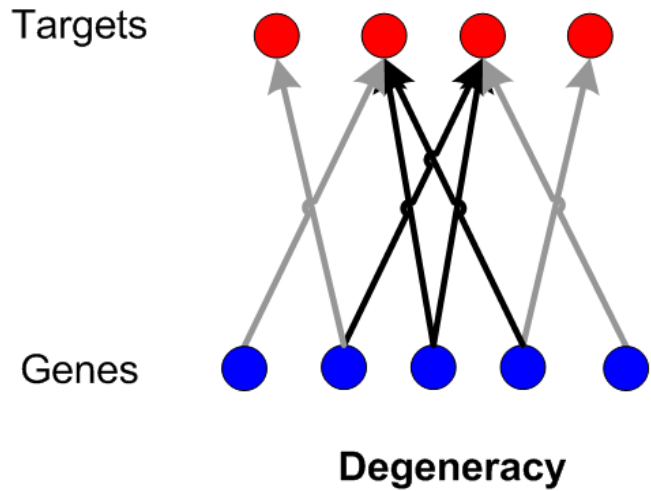

Figure 3

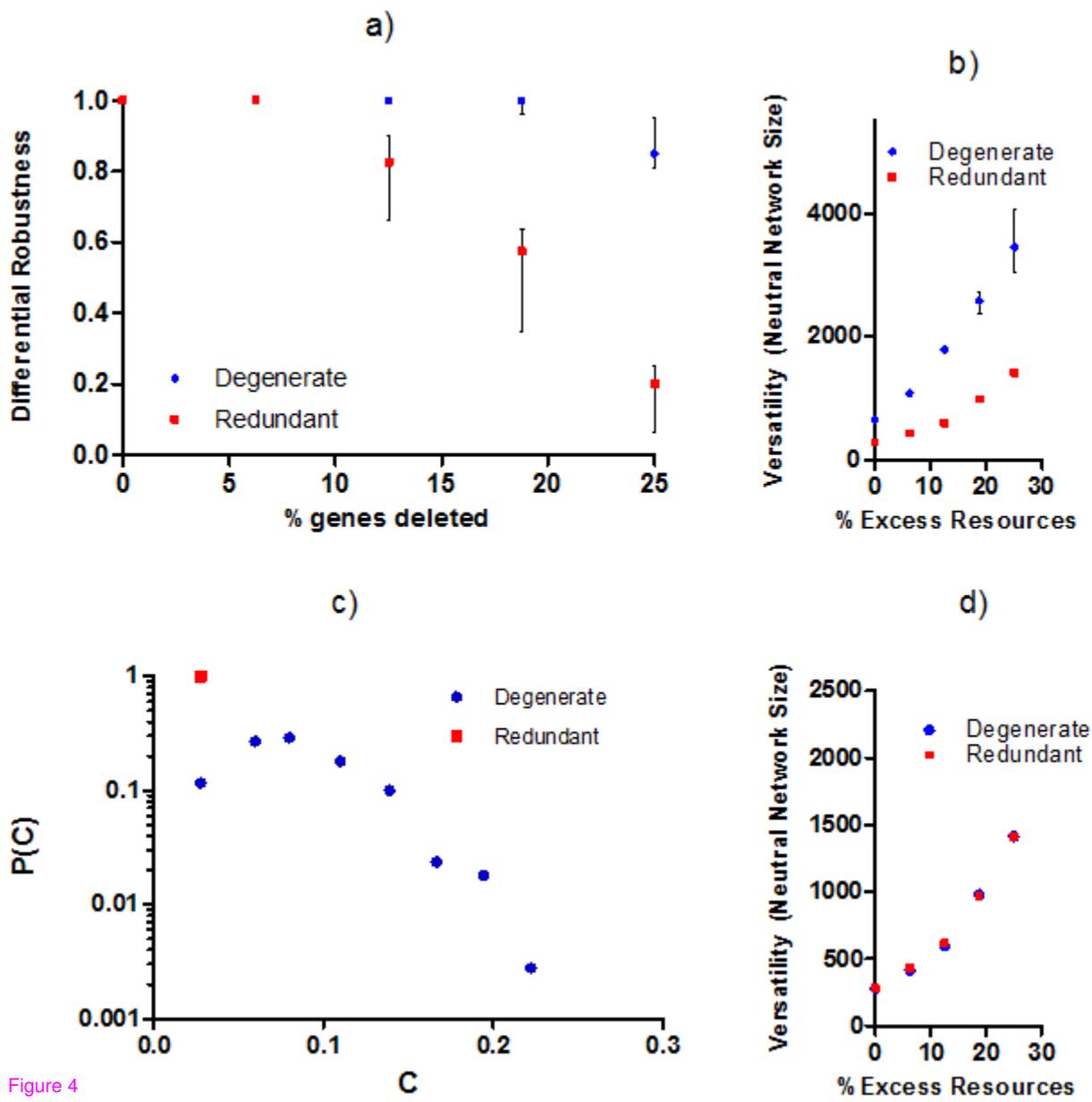

Figure 4

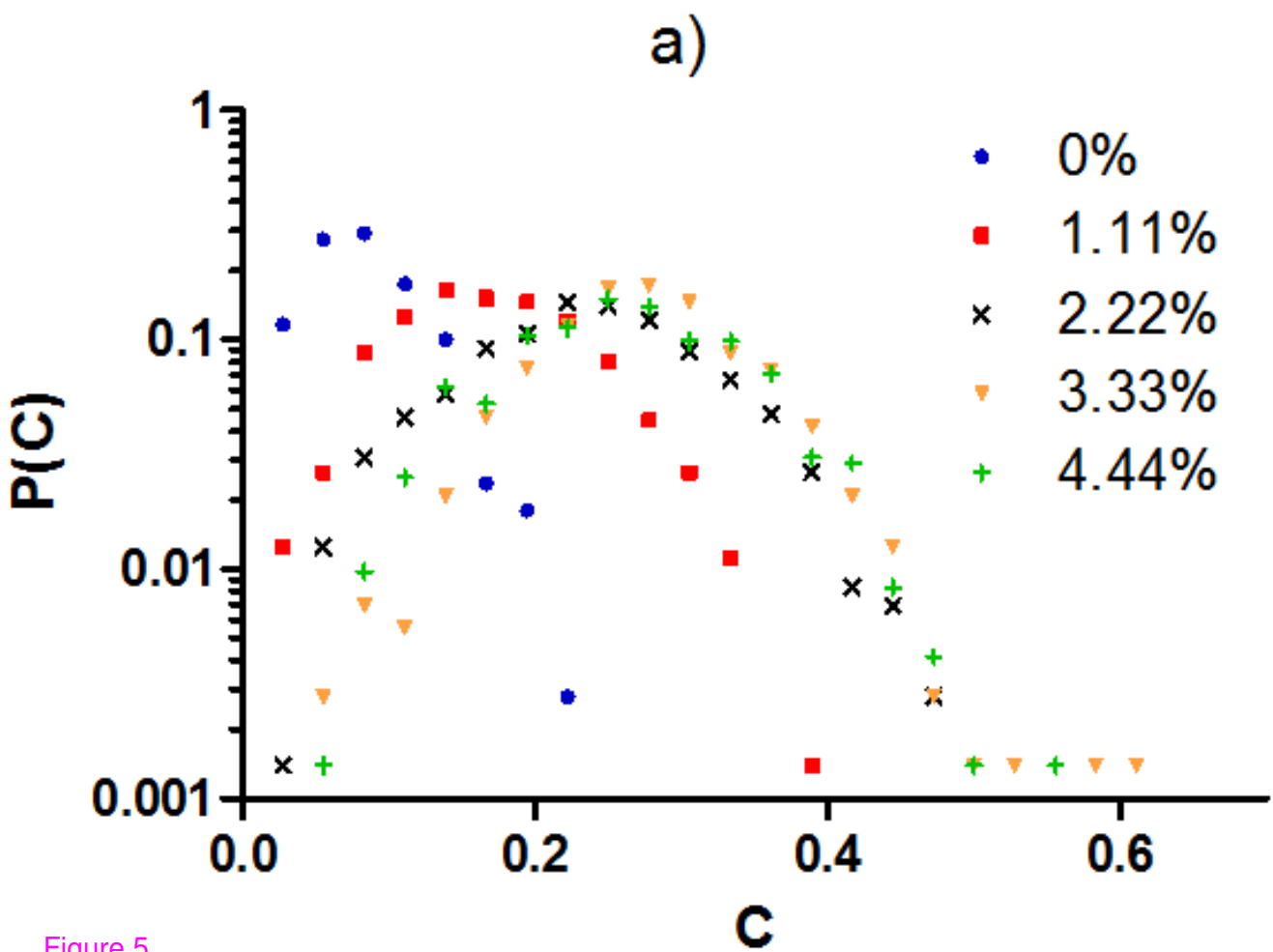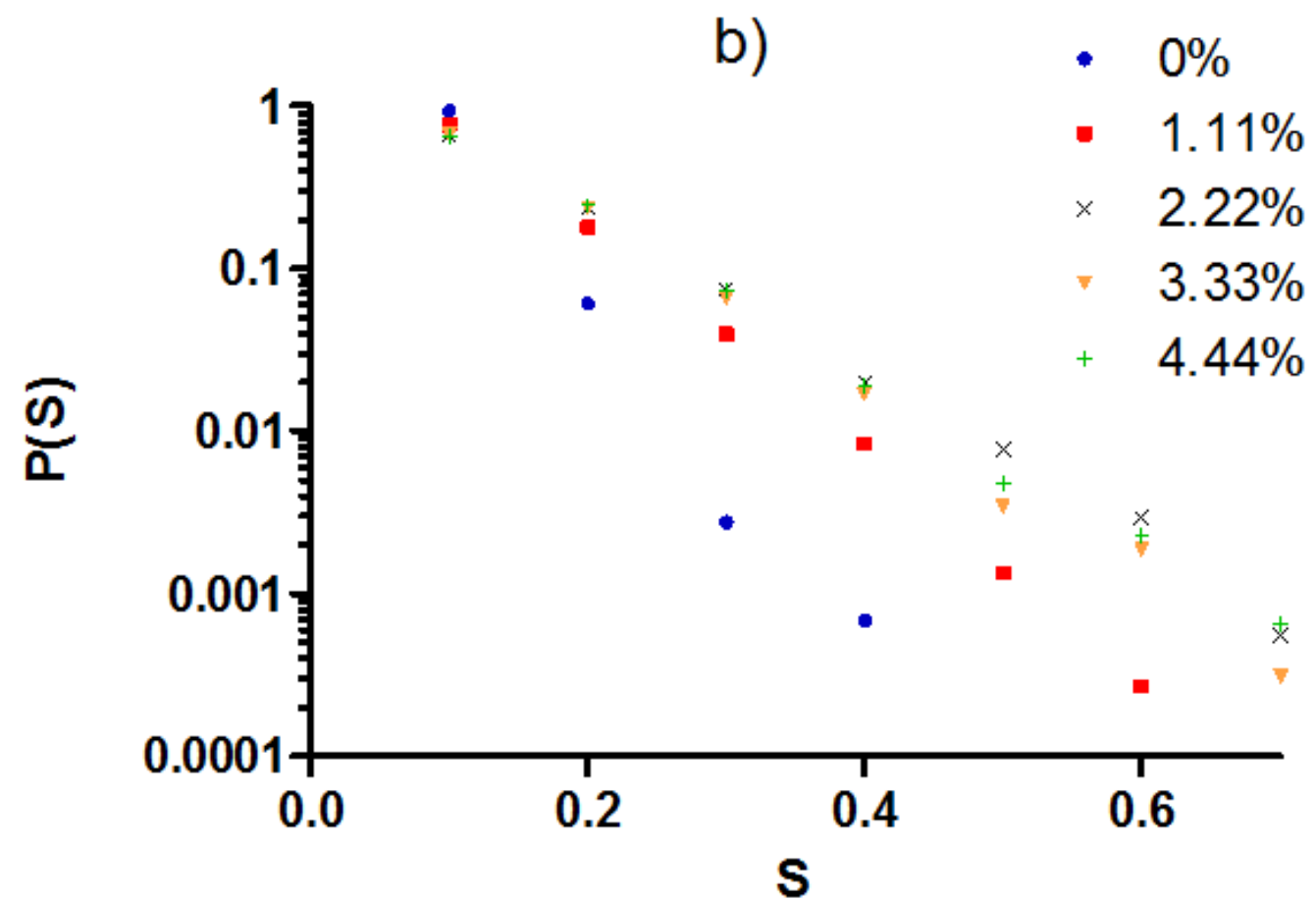

Figure 5